\documentclass[amsmath,amssymb,twocolumn]{revtex4}
\usepackage{graphicx}% Include figure files
\usepackage{amscd,amsmath,amsthm,amsfonts,amssymb}
\def\PT{$\cal{PT}$}
\def\[{\begin{equation}}
\def\]{\end{equation}}
\advance\textheight -16mm
\newtheorem{theorem}{Theorem}

\begin{document}
\title{General $N$-solitons and their dynamics in several nonlocal nonlinear Schr\"odinger equations}
\author{Jianke Yang}
\address{
{\small\it Department of Mathematics and Statistics, University of Vermont, Burlington, VT 05401, U.S.A} \\
{\normalsize \small \it Email: jyang@math.uvm.edu}}

\begin{abstract}
General $N$-solitons in three recently-proposed nonlocal nonlinear Schr\"odinger equations are presented. These nonlocal equations include the reverse-space, reverse-time, and reverse-space-time nonlinear Schr\"odinger equations, which are nonlocal reductions of the Ablowitz-Kaup-Newell-Segur (AKNS) hierarchy. It is shown that general $N$-solitons in these different equations can be derived from the same Riemann-Hilbert solutions of the AKNS hierarchy, except that symmetry relations on the scattering data are different for these equations. This Riemann-Hilbert framework allows us to identify new types of solitons with novel eigenvalue configurations in the spectral plane. Dynamics of $N$-solitons in these equations is also explored. In all the three nonlocal equations, a generic feature of their solutions is repeated collapsing. In addition, multi-solitons can behave very differently from fundamental solitons and may not correspond to a nonlinear superposition of fundamental solitons.
\end{abstract}

\maketitle

\section{Introduction}
Integrable systems have been studied for many years \cite{Ablowitz1981,Zakharov1984,Faddeev1987,Ablowitz1991,Yang2010}. Most such systems are local equations, i.e., the solution's evolution depends only on the local solution value and its local space and time derivatives. The Korteweg-de Vries equation and the nonlinear Schr\"odinger (NLS) equation are such examples.

A few years ago, a nonlocal reverse-space NLS equation
\[ \label{e:NLSRX}
iq_t(x,t)+q_{xx}(x,t)+2q^2(x,t)q^*(-x,t)=0,
\]
was proposed by Ablowitz and Musslimani \cite{AblowitzMussPRL2013}. Here the asterisk * represents complex conjugation. Although this equation is just a reduction of the Ablowitz-Kaup-Newell-Segur (AKNS) hierarchy, it is distinctive because the solution's evolution at location $x$ depends on not only the local solution at $x$, but also the nonlocal solution at the distant position $-x$. That is, solution states at distant locations $x$ and $-x$ are directly coupled, reminiscent of quantum entanglement between pairs of particles. Integrable equations of this type had not been paid attention before, which makes this nonlocal equation mathematically interesting. Regarding potential applications,
this nonlocal equation was linked to an unconventional system of magnetics \cite{PTNLSmagnetics}. In addition, since this equation is parity-time (\PT) symmetric, i.e.,  it is invariant under the joint transformations of $x\to -x$, $t\to -t$ and complex conjugation, it is thus related to the concept of \PT symmetry --- a hot research area of contemporary physics \cite{Yang_review}.

Following its introduction, this reverse-space NLS equation was actively studied \cite{AblowitzMussPRL2013,AblowitzMussNonli2016,WYY2016,HXLM2016,Gerdjikov2017,Stalin2017wrong,Caudrelier,Zhang1,Peckan2017arxiv,BYJYrogue}. In addition, many other nonlocal integrable equations were reported and investigated
\cite{AblowitzMussPRE2014,Yan,Khara2015,Fokas2016,GerdjikovNwave2016,Chow,JZN2017,Lou2,AblowitzMussSAPM,ZhoudNLS,
Zhu2,HePPTDS,Zhu3,BYJY2017,nonlocalFordy2017,AblowitzSG2017,BYnonlocalDS,Zhang2}.
These studies revealed interesting solution behaviors in nonlocal equations, such as finite-time solution blowup (i.e., collapsing) in fundamental solitons and general rogue waves of Eq. (\ref{e:NLSRX}) \cite{AblowitzMussPRL2013,BYJYrogue} and the simultaneous existence of solitons and kinks in the nonlocal modified Korteweg-de Vries equation \cite{Zhu2}. A connection between nonlocal and local equations was also discovered in \cite{BYJY2017}, where it was shown that many nonlocal equations could be converted to local equations through transformations.

In this article, we study general $N$-solitons and their dynamics in the reverse-space NLS equation (\ref{e:NLSRX}), as well as the
reverse-time and reverse-space-time NLS equations,
\[ \label{e:NLSRT}
iq_t(x,t)+q_{xx}(x,t)+2q^2(x,t)q(x,-t)=0,
\]
and
\[ \label{e:NLSRXT}
iq_t(x,t)+q_{xx}(x,t)+2q^2(x,t)q(-x,-t)=0.
\]
These equations can be derived from the following member of the AKNS hierarchy --- the coupled Schr\"odinger equations \cite{Ablowitz1981,Yang2010}
\[ \label{e:qr1}
iq_t+q_{xx}-2q^2r=0,
\]
\[ \label{e:qr2}
ir_t-r_{xx}+2 r^2q=0.
\]
Dynamics in these coupled Schr\"odinger equations without constraints between $q$ and $r$ has been analyzed in \cite{Yang2010}, and self-collapsing solitons as well as amplitude-changing solitons have been reported. Under reductions
\[ \label{e:reduce1}
r(x,t)=-q^*(-x, t),
\]
\[ \label{e:reduce2}
r(x,t)=-q(x, -t),
\]
and
\[ \label{e:reduce3}
r(x,t)=-q(-x, -t),
\]
these coupled Schr\"odinger equations reduce to the reverse-space NLS equation (\ref{e:NLSRX}),
reverse-time NLS equation (\ref{e:NLSRT}) and reverse-space-time NLS equation (\ref{e:NLSRXT})
respectively \cite{AblowitzMussPRL2013,AblowitzMussSAPM}.

This article is motivated by a number of reasons. First, while solitons in the reverse-space NLS equation (\ref{e:NLSRX}) have been investigated before \cite{AblowitzMussPRL2013,AblowitzMussNonli2016,HXLM2016,Stalin2017wrong}, only the fundamental solitons were reported \cite{AblowitzMussPRL2013,AblowitzMussNonli2016,HXLM2016}. In \cite{Stalin2017wrong}, both fundamental and two-solitons were also reported; but those solutions are clearly incorrect, as was pointed out in \cite{Peckan2017arxiv}. In \cite{WYY2016,Zhang1,Peckan2017arxiv},
``solitons" were also derived for Eq. (\ref{e:NLSRX}); however, those solutions are not true solitons since they are not localized in space.
Thus, despite the previous efforts, true multi-solitons in the reverse-space NLS equation (\ref{e:NLSRX}) have never been found, which is surprising. This motivates us to derive general multi-soliton solutions in this nonlocal equation. As we will show, multi-solitons in this equation admit novel eigenvalue configurations in the spectral space, which give rise to new types of soliton structures, such as the two-soliton in Fig. 2 (bottom row). In addition, multi-solitons behave very differently from fundamental solitons.

Our second motivation is that, there has been no studies of solitons in the reverse-time NLS equation (\ref{e:NLSRT}) and reverse-space-time NLS equation (\ref{e:NLSRXT}) to our best knowledge. The T-symmetric and ST-symmetric NLS equations studied in \cite{Peckan2017arxiv} and the reverse-$t$ NLS equation studied in \cite{Zhang2} are not the reverse-time and reverse-space-time NLS equations (\ref{e:NLSRT}) and (\ref{e:NLSRXT}). In fact, those equations are just the nonlocal nonlinear diffusion equations analyzed in \cite{BYJY2017}.

Our third motivation is that, it is helpful to put $N$-solitons of the three nonlocal equations (\ref{e:NLSRX})-(\ref{e:NLSRXT}) in the framework of inverse scattering and Riemann-Hilbert solutions, because in this framework, one can clearly see the novel symmetry relations in their scattering data, which strongly differ from those in the local (classical) NLS equation. In addition, this Riemann-Hilbert framework allows us to readily identify new types of solitons arising from new eigenvalue configurations in the spectral space, which can be more difficult to obtain by other methods (such as the Darboux transformation method and the bilinear method \cite{HXLM2016,Stalin2017wrong,Zhang1,Peckan2017arxiv,Zhang2}).

In this article, we derive general $N$-solitons in the reverse-space, reverse-time and reverse-space-time NLS equations (\ref{e:NLSRX})-(\ref{e:NLSRXT}) using the inverse scattering and Riemann-Hilbert method. We show that $N$-solitons in these different equations can be derived from the same Riemann-Hilbert solutions of the AKNS hierarchy, except that symmetry relations on their scattering data differ from each other (and from those of the local NLS equation). From this Riemann-Hilbert framework, we discover new types of multi-solitons with novel eigenvalue configurations in the spectral plane. Since these eigenvalue configurations cannot be split into groups of eigenvalues of fundamental solitons, we conclude that these multi-solitons cannot be viewed as nonlinear superpositions of fundamental solitons. Dynamics of these solitons is further analyzed. In all the three nonlocal equations, we show that a generic feature of their solitons is repeated collapsing. In addition, multi-solitons can behave very differently from fundamental solitons. For instance, in the reverse-time NLS equation (\ref{e:NLSRT}), a two-soliton can move in opposite directions and repeatedly collapse, while the fundamental soliton is always stationary and non-collapsing.

\section{$N$-solitons for general coupled Schr\"odinger equations}
Our basic idea to derive $N$-solitons in the reverse-space, reverse-time and reverse-space-time NLS equations (\ref{e:NLSRX})-(\ref{e:NLSRXT}) is to recognize that these equations are reductions of the coupled Schr\"odinger equations (\ref{e:qr1})-(\ref{e:qr2}). Thus, we will start with the Riemann-Hilbert solutions of $N$-solitons for these coupled Schr\"odinger equations for given scattering data, then impose appropriate symmetry relations on this scattering data, which will yield $N$-solitons for the underlying nonlocal equations. Following this approach, we first consider $N$-solitons for the coupled Schr\"odinger equations (\ref{e:qr1})-(\ref{e:qr2}), which will be done in this section.

The coupled Schr\"odinger equations (\ref{e:qr1})-(\ref{e:qr2}) are a member of the AKNS hierarchy, and their Lax pairs are \cite{ZS1972,AKNS1974}
\begin{equation} \label{ZS2}
 Y_x=MY, \quad Y_t=NY,
\end{equation}
where
\begin{equation}  \label{Yt2}
M=\left(\begin{array}{cc} -i\zeta & q \\ r &
 i\zeta\end{array}\right),   \quad
N=\left(\begin{array}{cc} -iqr-2i\zeta^2 & iq_x+2\zeta q \\ -ir_x+2\zeta r &
 iqr+2i\zeta^2 \end{array}\right).
\end{equation}
For localized functions $q(x,t)$ and $r(x,t)$, the inverse scattering transform was developed in \cite{ZS1972,AKNS1974}, and its modern Riemann-Hilbert treatment was developed in \cite{Zakharov1984,ZS1979}. Following this Riemann-Hilbert treatment, $N$-solitons in this system  were explicitly written down in \cite{Yang2010} as
 \[ \label{qsolution}
 q(x, t) = 2i\left(\sum\limits_{j,k=1}^N v_j \left(
 {M^{-1} }\right)_{jk}{{\bar v}_k}\right)_{12},
 \]
 and
 \[  \label{rsolution}
 r(x, t) = -2i\left(\sum\limits_{j,k=1}^N v_j \left(
 {M^{-1} }\right)_{jk}{{\bar v}_k}\right)_{21},
 \]
 where
 \[  \label{vk_no2}
 v_k(x, t)=e^{-i\zeta_k\Lambda x-2i\zeta_k^2\Lambda t} \hspace{0.03cm} v_{k0},
 \]
 \[  \label{vkbar_no2}
 \bar{v}_k(x, t)=\bar{v}_{k0} \hspace{0.04cm} e^{i\bar{\zeta}_k\Lambda x+2i\bar{\zeta}_k^2\Lambda t},
 \]
 $M$ is a $N\times N$ matrix whose $(j, k)$-th element is given by
\begin{equation} \label{Mdef_no2}
 M_{jk}={\frac{{\bar v}_j  v_k}{{\bar\zeta_j-\zeta_k }}}, \quad 1\le
 j, k \le N,
\end{equation}
$\Lambda=\mathrm{diag}(1,-1)$, $\zeta_k$ are complex numbers in the upper half plane $\mathbb{C}_+$, $\bar{\zeta}_k$ are complex numbers in the lower half plane $\mathbb{C}_-$, and $v_{k0}, \bar{v}_{k0}$ are constant column and row vectors of length two respectively.

The above solutions can be written in a more compact form. Let us denote
\[ \label{vk0form}
v_{k0}=\left[\begin{array}{c} a_k \\ b_k \end{array}\right],  \quad
\bar{v}_{k0} = \left[\bar{a}_k, \bar{b}_k\right],
\]
where $a_k, b_k, \bar{a}_k, \bar{b}_k$ are complex constants, and
\[
\theta_k=-i\zeta_k x-2i\zeta_k^2 t, \quad \bar{\theta}_k=i\bar{\zeta}_k x+2i\bar{\zeta}_k^2 t.
\]
Notice that $M^{-1}$ in solutions (\ref{qsolution})-(\ref{rsolution}) can be expressed as the transpose of $M$'s
cofactor matrix divided by $\det M$. Also recall that the determinant of a matrix can be expressed as the sum of its elements
along a row or column multiplying their corresponding cofactors. Hence solutions (\ref{qsolution})-(\ref{rsolution}) can be rewritten as ratios of determinants \cite{Faddeev1987,Yang2010}
\[ \label{Nsoliton}
q(x, t)=-2i \frac{\det F}{\det M}, \quad r(x, t)=2i \frac{\det G}{\det M},
\]
where $F$ and $G$ are the following $(N+1)\times (N+1)$ matrices:
\[  \label{FNLS}
F=\left(\begin{array}{cccc} 0 & a_1 e^{\theta_1} &  \dots & a_N e^{\theta_N} \\
\bar{b}_1 e^{-\bar{\theta}_1} & M_{11} & \dots & M_{1N} \\
 \vdots & \vdots & \vdots & \vdots \\
\bar{b}_N e^{-\bar{\theta}_N} & M_{N1} & \dots & M_{NN} \end{array}\right),
\]
and
\[  \label{GNLS}
G=\left(\begin{array}{cccc} 0 & b_1 e^{-\theta_1} &  \dots & b_N e^{-\theta_N} \\
\bar{a}_1 e^{\bar{\theta}_1} & M_{11} & \dots & M_{1N} \\
 \vdots & \vdots & \vdots & \vdots \\
\bar{a}_N e^{\bar{\theta}_N} & M_{N1} & \dots & M_{NN} \end{array}\right).
\]

The reverse-space, reverse-time and reverse-space-time NLS equations (\ref{e:NLSRX})-(\ref{e:NLSRXT}) were obtained from the coupled system (\ref{e:qr1})-(\ref{e:qr2}) under reductions (\ref{e:reduce1})-(\ref{e:reduce3}). Each reduction leads to its own
symmetry relations on the discrete scattering data $\{\zeta_k, \bar{\zeta}_k, v_{k0}, \bar{v}_{k0}, 1\le k\le N\}$. For reasons which will become apparent in the next section, we call $\zeta_k, \bar{\zeta}_k$ eigenvalues and $v_{k0}, \bar{v}_{k0}$ eigenvectors in this paper. By deriving these symmetry relations for the eigenvalues and eigenvectors of the scattering data, $N$-soliton solutions of nonlocal equations (\ref{e:NLSRX})-(\ref{e:NLSRXT}) will be obtained directly from the above general $N$-soliton formulae (\ref{Nsoliton}). This will be done in the next section.

\section{Symmetry relations of scattering data in the nonlocal NLS equations}

We first present symmetry relations of the scattering data for the reverse-space, reverse-time and reverse-space-time nonlocal NLS equations (\ref{e:NLSRX})-(\ref{e:NLSRXT}), followed by their proofs. For this purpose, we introduce some notations. We define
\[
\sigma_1=\left[\begin{array}{cc} 0 & 1 \\ 1 & 0\end{array}\right], \nonumber
\]
which is a Pauli spin matrix, and use the superscript `$T$' to represent the transpose of a matrix. In addition, we use $\mathbb{R}_{\pm}$ to represent the sets of positive and negative real numbers respectively.
\begin{theorem} \label{Theorem1}
For the reverse-space NLS equation (\ref{e:NLSRX}), if $\zeta$ is an eigenvalue, so is $-\zeta^*$. Thus, non-purely-imaginary eigenvalues appear as pairs $(\zeta, -\zeta^*)$, which lie in the same half of the complex plane. Symmetry relations on the eigenvectors are given as follows.
\begin{enumerate}
\item For a pair of non-purely-imaginary eigenvalues $(\zeta_k, \hat{\zeta}_k) \in \mathbb{C}_+$,
$(\zeta_k, \hat{\zeta}_k)\notin i\mathbb{R}_+$, with $\hat{\zeta}_k=-\zeta^*_k$, their column eigenvectors $v_{k0}$ and $\hat{v}_{k0}$ are related as $\hat{v}_{k0}=\sigma_1 v_{k0}^*$.
\item For a purely imaginary eigenvalue $\zeta_k$ $\in i\mathbb{R}_+$, its eigenvector is of the form $v_{k0}=\left[1, \, e^{i\theta_k}\right]^T$, where $\theta_k$ is a real constant.
\item For a pair of non-purely-imaginary eigenvalues $(\bar{\zeta}_k, \hat{\bar{\zeta}}_k) \in \mathbb{C}_-$, $(\bar{\zeta}_k, \hat{\bar{\zeta}}_k) \notin i\mathbb{R}_-$, with $\hat{\bar{\zeta}}_k=-\bar{\zeta}^*_k$, their row eigenvectors $\bar{v}_{k0}$ and $\hat{\bar{v}}_{k0}$ are related as $\hat{\bar{v}}_{k0}= \bar{v}_{k0}^*\sigma_1$.
\item For a purely imaginary eigenvalue $\bar{\zeta}_k$ $\in i\mathbb{R}_-$, its eigenvector is of the form $\bar{v}_{k0}=\left[1, \, e^{i\bar{\theta}_k}\right]$, where $\bar{\theta}_k$ is a real constant.
\end{enumerate}
\end{theorem}

\begin{theorem} \label{Theorem2}
For the reverse-time NLS equation (\ref{e:NLSRT}), if $\zeta$ is an eigenvalue, so is $-\zeta$. Thus, eigenvalues appear as pairs $(\zeta, -\zeta)$,  which lie on the opposite halves of the complex plane. For a pair of such eigenvalues $(\zeta_k, \bar{\zeta}_k)$ with $\zeta_k \in \mathbb{C}_+$ and $\bar{\zeta}_k=-\zeta_k \in \mathbb{C}_-$, their eigenvectors $v_{k0}$ and $\bar{v}_{k0}$ are related as $\bar{v}_{k0}=v_{k0}^T$.
\end{theorem}

\begin{theorem} \label{Theorem3}
For the reverse-space-time NLS equation (\ref{e:NLSRXT}), eigenvalues $\zeta_k$ can be anywhere in $\mathbb{C}_+$, and eigenvalues $\bar{\zeta}_k$ can be anywhere in $\mathbb{C}_-$. However, their eigenvectors must be of the forms
\[
v_{k0}=\left[1, \omega_k\right]^T, \quad  \bar{v}_{k0}=\left[1, \bar{\omega}_k\right],  \nonumber
\]
where $\omega_k=\pm 1$, and $\bar{\omega}_k=\pm 1$.
\end{theorem}

To put these results in perspective, we recall that for the local NLS equation, which is obtained from the coupled Schr\"odinger equations (\ref{e:qr1})-(\ref{e:qr2}) under the reduction of $r(x,t)=-q^*(x, t)$, the symmetries of its scattering data are $\bar{\zeta}_k=\zeta_k^*$ and $\bar{v}_{k0}=v_{k0}^{*T}$ \cite{Ablowitz1981,Zakharov1984,Yang2010}. Thus,
symmetry relations for the nonlocal NLS equations are very different from those of the local NLS equation.  In particular, for the reverse-space and reverse-space-time NLS equations, eigenvalues in the upper and lower halves of the complex plane are completely independent. This independence allows for novel eigenvalue configurations, which will give rise to new types of multi-solitons. This will be demonstrated in the next section.

Before proving these theorems, we first establish a connection between the discrete scattering data for $N$-solitons, $\{\zeta_k, \bar{\zeta}_k, a_k, b_k, \bar{a}_k, \bar{b}_k, 1\le k\le N\}$, and discrete eigenmodes in the eigenvalue problem $Y_x=MY$ and its adjoint eigenvalue problem $K_x=-KM$,
i.e.,
\[ \label{e:Yx}
Y_x=-i\zeta \Lambda Y +QY,
\]
and
\[ \label{e:Kx}
K_x=i\zeta K\Lambda-KQ,
\]
where the potential matrix $Q$ is
\[
Q(x)=\left[\begin{array}{cc} 0 & q(x, 0) \\ r(x,0) & 0 \end{array}\right],
\]
and $q(x,0), r(x,0)$ are the initial conditions of functions $q(x,t)$ and $r(x,t)$.
Indeed, it is known, from \cite{Yang2010} for instance, that each subset $\{\zeta_k, a_k, b_k\}$ of the discrete scattering data, with $\zeta_k\in \mathbb{C}_+$, corresponds to a discrete eigenvalue $\zeta_k$ in the eigenvalue problem (\ref{e:Yx}), whose discrete eigenfunction $Y_k(x)$ has the following asymptotics
\[ \label{Yasym1a}
Y_k(x) \longrightarrow \left[\begin{array}{c} a_k e^{-i\zeta_k x} \\ 0 \end{array} \right], \quad x \to -\infty,
\]
\[  \label{Yasym1b}
Y_k(x) \longrightarrow \left[\begin{array}{c} 0 \\ -b_k e^{i\zeta_k x} \end{array} \right], \quad x \to +\infty.
\]
Analogously, each subset $\{\bar{\zeta}_k, \bar{a}_k, \bar{b}_k\}$ of the discrete scattering data, with $\bar{\zeta}_k\in \mathbb{C}_-$, corresponds to a discrete eigenvalue $\bar{\zeta}_k$ in the adjoint eigenvalue problem (\ref{e:Kx}), whose discrete eigenfunction $K_k(x)$ has the following asymptotics
\[  \label{Kasym1a}
K_k(x) \longrightarrow \left[\bar{a}_k e^{i\bar{\zeta}_k x}, \quad 0 \right], \qquad x \to -\infty,
\]
\[  \label{Kasym1b}
K_k(x) \longrightarrow \left[0, \quad -\bar{b}_k e^{-i\bar{\zeta}_k x} \right], \quad x \to +\infty.
\]
In view of this connection, in order to derive symmetry relations on the (discrete) scattering data, we will use symmetry relations of discrete eigenmodes in the eigenvalue problems (\ref{e:Yx})-(\ref{e:Kx}), as we will do below. This symmetry derivation is easier than the standard one in \cite{Ablowitz1981,Zakharov1984,Yang2010}, because it does not use details of the scattering theory for the underlying integrable equations.

\vspace{0.15cm}
\textbf{Proof of Theorem \ref{Theorem1}.}
The reverse-space NLS equation (\ref{e:NLSRX}) was derived from the coupled Schr\"odinger equations (\ref{e:qr1})-(\ref{e:qr2}) under the reduction (\ref{e:reduce1}). With this reduction, the potential matrix $Q$ is
\[
Q(x)=\left[\begin{array}{cc} 0 & q(x,0) \\ -q^*(-x, 0) & 0\end{array}\right],
\]
which features the following symmetry,
\[
Q^*(-x)=-\sigma_1^{-1}Q(x)\sigma_1.
\]
Taking the complex conjugate to the eigenvalue equation (\ref{e:Yx}), reversing $x$ to $-x$, and utilizing the above potential symmetry, we get
\[
\widehat{Y}_x=-i\hat{\zeta} \Lambda\widehat{Y}+Q\widehat{Y},
\]
where
\[ \label{e:Yxalpha}
\hat{\zeta}=-\zeta^*, \quad \widehat{Y}(x)=\alpha\sigma_1Y^*(-x),
\]
and $\alpha$ is an arbitrary complex constant.
This equation shows that, if $\zeta_k \in \mathbb{C}_+$ is an eigenvalue of the scattering problem (\ref{e:Yx}), so is $\hat{\zeta}_k\equiv -\zeta^*_k \in \mathbb{C}_+$. In addition, the eigenfunction $Y_k(x)$ of $\zeta_k$ and the eigenfunction $\widehat{Y}_k(x)$ of $\hat{\zeta}_k$ are related as in (\ref{e:Yxalpha}). Recall that the large-$x$ asymptotics of $\zeta_k$'s eigenfunction $Y_k(x)$ has been given in Eqs. (\ref{Yasym1a})-(\ref{Yasym1b}), and the large-$x$ asymptotics of $\hat{\zeta}_k$'s eigenfunction $\hat{Y}_k(x)$ is
\[ \label{Yasym2a}
\hat{Y}_k(x) \longrightarrow \left[\begin{array}{c} \hat{a}_k e^{-i\hat{\zeta}_k x} \\ 0 \end{array} \right], \quad x \to -\infty.
\]
\[  \label{Yasym2b}
\hat{Y}_k(x) \longrightarrow \left[\begin{array}{c} 0 \\ -\hat{b}_k e^{i\hat{\zeta}_k x} \end{array} \right], \quad x \to +\infty,
\]
where $\hat{a}_k$ and $\hat{b}_k$ are complex constants. Utilizing these asymptotics, the eigenfunction relation in Eq. (\ref{e:Yxalpha}) reveals that
\[
\hat{a}_k=-\alpha b_k^*, \quad \hat{b}_k=-\alpha a_k^*,
\]
i.e.,
\[ \label{e:valpha}
\hat{v}_{k0}=-\alpha \sigma_1 v_{k0}^*.
\]

If $\zeta_k$ is not purely imaginary, then its counterpart $\hat{\zeta}_k= -\zeta^*_k$ is a different eigenvalue. In this case, when the above $\hat{v}_{k0}$ expression is inserted into the $N$-soliton formulae (\ref{Nsoliton}), the constant $-\alpha$ cancels out and does not contribute to the solution. Thus, we can set $-\alpha=1$ without loss of generality. Then, $\hat{v}_{k0}=\sigma_1 v_{k0}^*$, and part 1 of Theorem \ref{Theorem1} is proved.

If $\zeta_k$ is purely imaginary, then $\hat{\zeta}_k=\zeta_k$. Thus, their eigenvectors are also the same, i.e., $\hat{v}_{k0}=v_{k0}$. Without loss of generality, we can scale the eigenvector $v_{k0}$ so that its first element $a_k=1$. Then, inserting $\hat{v}_{k0}=v_{k0}$ into Eq.~(\ref{e:valpha}), we find that $|\alpha|=1$ and $v_{k0}=[1, -\alpha]^T$. Denoting $-\alpha=e^{i\theta_k}$, where $\theta_k$ is a real constant, we get $v_{k0}=[1, e^{i\theta_k}]^T$; hence part 2 of Theorem \ref{Theorem1} is proved.

Repeating the above arguments on the adjoint eigenvalue problem (\ref{e:Kx}), parts 3 and 4 of Theorem \ref{Theorem1} can be similarly proved. $\Box$

\vspace{0.15cm}
\textbf{Proof of Theorem \ref{Theorem2}.}
The reverse-time NLS equation (\ref{e:NLSRT}) was derived from the coupled Schr\"odinger equations (\ref{e:qr1})-(\ref{e:qr2}) under the reduction (\ref{e:reduce2}). With this reduction, the potential matrix $Q$ is
\[
Q(x)=\left[\begin{array}{cc} 0 & q(x,0) \\ -q(x, 0) & 0\end{array}\right],
\]
which features the following symmetry,
\[
Q^T(x)=-Q(x).
\]
Then, taking the transpose of the eigenvalue problem (\ref{e:Yx}) and utilizing the above potential symmetry, we get
\[ \label{Ybar}
\overline{Y}_x=i\bar{\zeta}\hspace{0.07cm} \overline{Y} \Lambda- \overline{Y} Q,
\]
where
\[ \label{Ybar2}
\bar{\zeta}=-\zeta, \quad \overline{Y}(x)=Y^T(x).
\]
Eq. (\ref{Ybar}) means that $\left[\bar{\zeta}, \overline{Y}(x)\right]$ satisfies the adjoint eigenvalue equation (\ref{e:Kx}). Thus,
if $\zeta_k\in \mathbb{C}_+$ is an eigenvalue of the scattering problem (\ref{e:Yx}), then $\bar{\zeta}_k=-\zeta_k \in \mathbb{C}_-$ is an eigenvalue of the adjoint scattering problem (\ref{e:Kx}), and their eigenfunctions are related as in Eq. (\ref{Ybar2}). Utilizing this eigenfunction relation as well as the large-$x$ asymptotics of the eigenfunctions and adjoint eigenfunctions in Eqs. (\ref{Yasym1a})-(\ref{Kasym1b}), we readily find that $\bar{a}_k=a_k$ and $\bar{b}_k=b_k$, i.e., $\bar{v}_{k0}=v_{k0}^T$. Theorem \ref{Theorem2} is then proved. $\Box$

\vspace{0.15cm}
\textbf{Proof of Theorem \ref{Theorem3}.}
The reverse-space-time NLS equation (\ref{e:NLSRXT}) was derived from the coupled Schr\"odinger equations (\ref{e:qr1})-(\ref{e:qr2}) under the reduction (\ref{e:reduce3}). With this reduction, the potential matrix $Q$ is
\[
Q(x)=\left[\begin{array}{cc} 0 & q(x,0) \\ -q(-x, 0) & 0\end{array}\right],
\]
which features the symmetry,
\[
Q(-x)=-\sigma_1^{-1} Q(x)\sigma_1.
\]
Reversing $x$ to $-x$ in the eigenvalue problem (\ref{e:Yx}) and utilizing the above potential symmetry, we get
\[
\widehat{Y}_x=-i\zeta \Lambda\widehat{Y}+Q\widehat{Y},
\]
where
\[ \label{e:Yx3}
\widehat{Y}(x)=\sigma_1 Y(-x).
\]
This equation means that for any eigenvalue $\zeta_k\in \mathbb{C}_+$, if $Y_k(x)$ is its eigenfunction, so is $\widehat{Y}_k(x)=\sigma_1 Y(-x)$; thus $\widehat{Y}_k(x)$ and $Y_k(x)$ are linearly dependent, i.e.,
\[
Y_k(x)=-\omega_k \sigma_1 Y_k(-x),
\]
where $\omega_k$ is some constant. Utilizing this relation and the large-$x$ asymptotics of the eigenfunction $Y_k(x)$ in Eqs. (\ref{Yasym1a})-(\ref{Yasym1b}), we find that $a_k=\omega_k b_k$ and $b_k=\omega_k a_k$; thus $\omega_k=\pm 1$. Without loss of generality, we scale the eigenvector $v_{k0}$ so that $a_k=1$. Then, $b_k=\omega_k$, and $v_{k0}=[1, \omega_k]^T$.

Since Eq. (\ref{e:Yx3}) also means that for any eigenvalue $\bar{\zeta}_k \in \mathbb{C}_-$, if $K_k(x)$ is its adjoint eigenfunction, so is $\widehat{K}_k(x)=\sigma_1 K_k(-x)$. Hence utilizing this relation and the large-$x$ asymptotics of the adjoint eigenfunction $K_k(x)$ in Eqs. (\ref{Kasym1a})-(\ref{Kasym1b}), we can similarly show that $\bar{v}_{k0}=[1, \bar{\omega}_k]$, where $\bar{\omega}_k=\pm 1$. Theorem \ref{Theorem3} is then proved. $\Box$

Before concluding this section, we point that it is also possible to impose $(q, r)$ reductions (\ref{e:reduce1})-(\ref{e:reduce3}) directly on the determinant solutions (\ref{Nsoliton}) in order to extract symmetry relations on the scattering data $\{\zeta_k, \bar{\zeta}_k, v_{k0}, \bar{v}_{k0}, 1\le k\le N\}$. However, our derivation of these relations above is easier. In addition, this derivation is more insightful since it is in the inverse-scattering and Riemann-Hilbert framework.

\section{Dynamics of $N$-solitons in the reverse-space NLS equation}
To obtain general $N$-solitons in the reverse-space NLS equation (\ref{e:NLSRX}), we only need to substitute the symmetry relations of the discrete scattering data in Theorem \ref{Theorem1} into the general $N$-soliton formulae (\ref{Nsoliton}). The only thing we want to add is that, for a pair of non-imaginary eigenvalues $(\zeta_k, -\zeta_k^*)\in \mathbb{C}_+$,
since we can scale the eigenvector $v_{k0}$ of $\zeta_k$ so that $a_k=1$ in Eq.~(\ref{vk0form}), then using Theorem \ref{Theorem1}, we get
\[
v_{k0}=[1, b_k]^T, \quad \hat{v}_{k0}=[b_k^*, 1]^T,
\]
where $\hat{v}_{k0}$ is the eigenvector of eigenvalue $\hat{\zeta}_k\equiv -\zeta_k^*$, and $b_k$ is a complex constant. Similarly, for a pair of non-imaginary eigenvalues $(\bar{\zeta}_k, -\bar{\zeta}_k^*) \in \mathbb{C}_-$, we can set their eigenvectors as
\[
\bar{v}_{k0}=[1, \bar{b}_k], \quad \hat{\bar{v}}_{k0}=[\bar{b}_k^*, 1],
\]
where $\bar{b}_k$ is another complex constant.

\subsection{Fundamental solitons}
First, we consider the fundamental (simplest) soliton solutions. These solutions correspond to a single pair of purely imaginary eigenvalues,
$\zeta_1=i\eta_1\in i\mathbb{R}_+$, and $\bar{\zeta}_1=i\bar{\eta}_1\in i\mathbb{R}_-$, where $\eta_1>0$ and $\bar{\eta}_1<0$. Eigenvectors $v_{10}$ and $\bar{v}_{10}$ of these eigenvalues are as given in Theorem \ref{Theorem1}, i.e., $v_{10}=\left[1, e^{i\theta_1}\right]^T$, and $\bar{v}_{10}=\left[1, \, e^{i\bar{\theta}_1}\right]$, where $\theta_1, \bar{\theta}_1$ are real constants. Substituting these expressions into the $N$-soliton formulae (\ref{Nsoliton}), we obtain the expression for the fundamental soliton in the reverse-space NLS equation (\ref{e:NLSRX}) as
\[ \label{1solitonRX}
q(x,t)=\frac{2(\eta_1-\bar{\eta}_1)e^{2\bar{\eta}_1x+4i\bar{\eta}_1^2t+i\bar{\theta}_1}}
{1+e^{-2(\eta_1-\bar{\eta}_1)x-4i(\eta_1^2-\bar{\eta}_1^2)t+i(\theta_1+\bar{\theta}_1)}},
\]
which agrees with that derived in \cite{AblowitzMussPRL2013,AblowitzMussNonli2016}.
This soliton has four free real parameters, $\eta_1, \bar{\eta}_1, \theta_1$ and $\bar{\theta}_1$ --- the same number of free parameters as the fundamental soliton in the local NLS equation. However, the present soliton can not move in space regardless of the choice of parameter values, which contrasts that in the local NLS equation.
Another general feature of this solution is that, if $\bar{\eta}_1 \ne -\eta_1$, i.e., $\bar{\zeta}_1 \ne -\zeta_1$,
then it would breathe and periodically collapse in time at position $x=0$. The period of this collapse is $\pi/[2(\eta_1^2-\bar{\eta}_1^2)]$.
To illustrate, we take parameter values
\[ \label{1solitonRXpar}
\eta_1=1, \quad \bar{\eta}_1=-0.5, \quad \theta_1=\pi/4, \quad \bar{\theta}_1=0.
\]
In this case, the locations of eigenvalues $\zeta_1$ and $\bar{\zeta}_1$ are shown in the left panel of Fig. 1, and the graph of the corresponding fundamental soliton is shown in the right panel of this figure.

\begin{figure}[htbp]
\begin{center}
\includegraphics[width=0.5\textwidth]{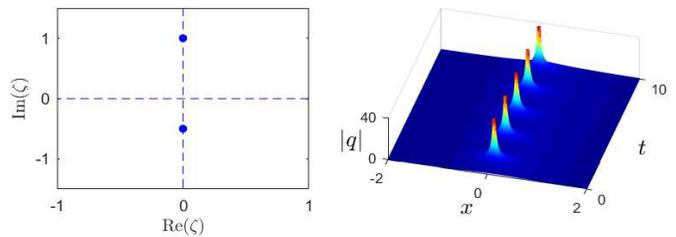}
\caption{A fundamental soliton (\ref{1solitonRX}) in the reverse-space NLS equation (\ref{e:NLSRX}) with parameters (\ref{1solitonRXpar}). Left panel: locations of eigenvalues in the complex plane. Right panel: solution graph. } \label{f:fig1}
\end{center}
\end{figure}

If $\bar{\eta}_1 = -\eta_1$, i.e., $\bar{\zeta}_1 =-\zeta_1$, then as long as $\theta_1+\bar{\theta}_1\ne (2n+1)\pi$ for any integer $n$,
this soliton will not collapse, and its amplitude $|q(x,t)|$ will not change with time \cite{AblowitzMussPRL2013,AblowitzMussNonli2016}.

\subsection{Two-solitons}
Now we consider two-solitons, which correspond to four eigenvalues, with $\zeta_1, \zeta_2 \in \mathbb{C}_+$ and $\bar{\zeta}_1, \bar{\zeta}_2 \in \mathbb{C}_-$. From Theorem \ref{Theorem1}, we see that $(\zeta_1, \zeta_2)$ in $\mathbb{C}_+$ and $(\bar{\zeta}_1, \bar{\zeta}_2)$ in $\mathbb{C}_-$ are totally independent. Thus, these four eigenvalues can be arranged in 4 different configurations.

\vspace{0.2cm}
\noindent
(1) $\zeta_1, \zeta_2 \in i\mathbb{R}_+$, and $\bar{\zeta}_1, \bar{\zeta}_2 \in i\mathbb{R}_-$.

\begin{figure}[htbp]
\begin{center}
\vspace{0.5cm}
\includegraphics[width=0.5\textwidth]{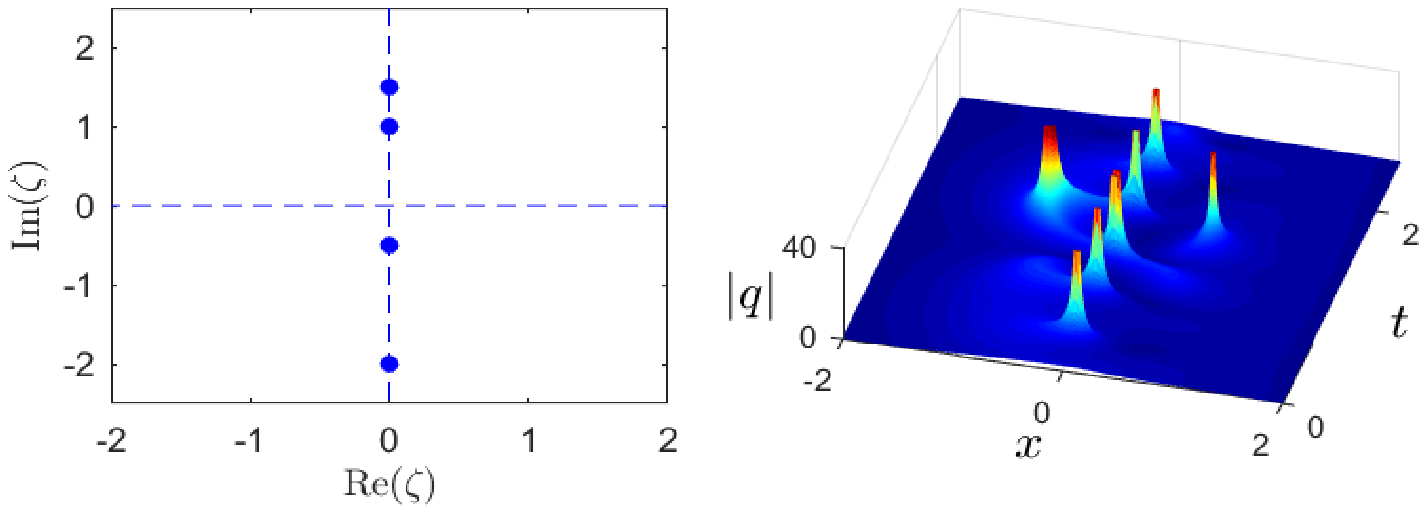}

\includegraphics[width=0.5\textwidth]{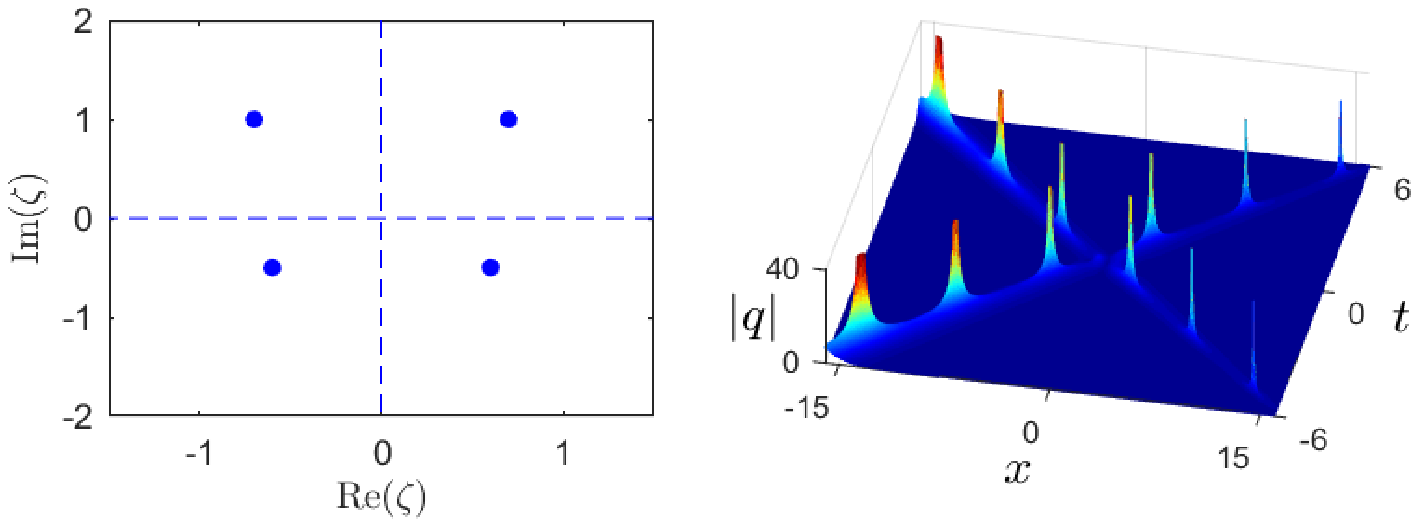}

\includegraphics[width=0.5\textwidth]{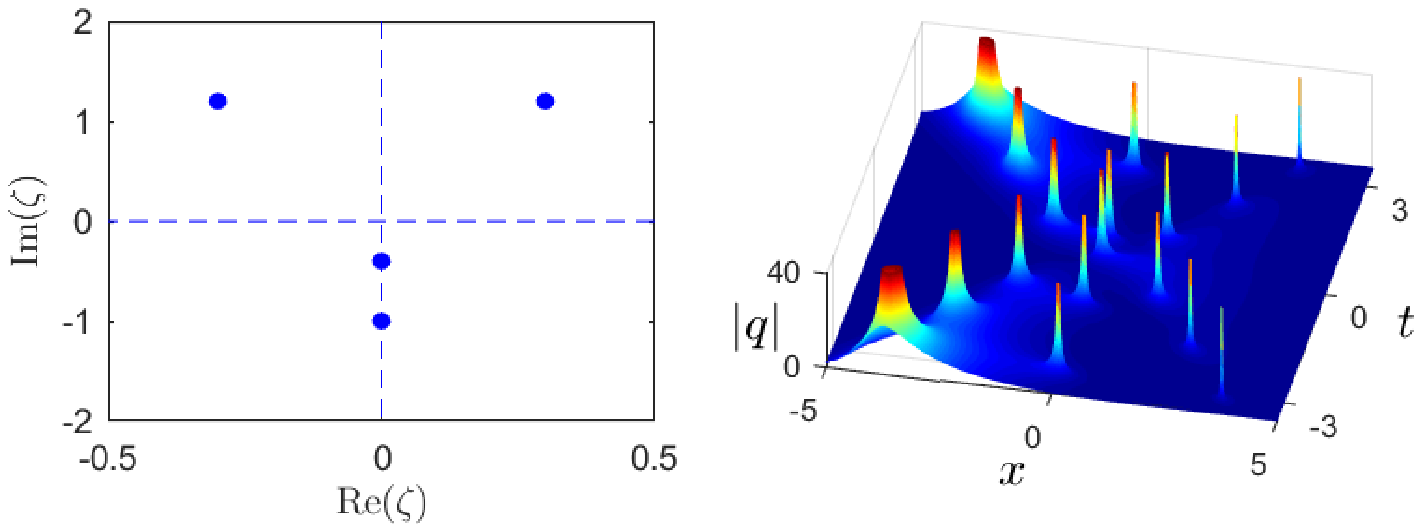}
\caption{Three examples of two-solitons in the reverse-space NLS equation (\ref{e:NLSRX}). Parameters for these solitons (from top to bottom) are given in equations (\ref{2solitonRXpar1a})-(\ref{2solitonRXpar1b}), (\ref{2solitonRXpar2a})-(\ref{2solitonRXpar2b}), and (\ref{2solitonRXpar3a})-(\ref{2solitonRXpar3b}), respectively. Left columns: eigenvalue configurations; right columns: solution graphs. } \label{f:fig2}
\end{center}
\end{figure}

\vspace{0.1cm}
In this case, all four eigenvalues are purely imaginary. Thus, the two-soliton solution is obtained from the $N$-soliton formula (\ref{Nsoliton}) with $N=2$, and the eigenvectors are given by Theorem \ref{Theorem1} as
\[
v_{k0}=\left[1, e^{i\theta_k}\right]^T, \quad \bar{v}_{k0}=\left[1, \, e^{i\bar{\theta}_k}\right],  \quad k=1, 2,
\]
where $\theta_1, \theta_2, \bar{\theta}_1$ and $\bar{\theta}_2$ are free real constants. Together with the four free real constants in the eigenvalues, this two-soliton has 8 free real parameters. We find that this soliton does not move, similar to the fundamental soliton. In addition, if $\bar{\zeta}_1 \ne -\zeta_1$ or $\bar{\zeta}_2 \ne -\zeta_2$, then it would repeatedly collapse, mostly at $x=0$, but occasionally at pairs of other spatial locations symmetric with respect to $x=0$ as well. An example is shown in Fig. \ref{f:fig2} (top row), where the parameters are chosen as
\begin{eqnarray}
&& \hspace{-0.5cm}
\zeta_1=i, \quad \zeta_2=1.5i, \quad \bar{\zeta}_1=-0.5i, \quad \bar{\zeta}_2=-2i,     \label{2solitonRXpar1a}  \\
&& \hspace{-0.5cm}
\theta_1=\pi/4, \quad \theta_2=0, \quad \bar{\theta}_1=0, \quad \bar{\theta}_2=\pi/2.  \label{2solitonRXpar1b}
\end{eqnarray}

If $\bar{\zeta}_1 = -\zeta_1$ and $\bar{\zeta}_2 = -\zeta_2$, then this solution could be bounded for all space and time, or collapse repeatedly, depending on the parameter values of $\theta_1, \theta_2, \bar{\theta}_1, \bar{\theta}_2$. For instance, if
$\zeta_1=-\bar{\zeta}_1=i$ and $\zeta_2=-\bar{\zeta}_2=2i$, then this solution is bounded when $\theta_1=0, \theta_2=\pi/2, \bar{\theta}_1=0, \bar{\theta}_2=-\pi/2$, but repeatedly collapse when $\theta_1=\pi/4, \theta_2=1, \bar{\theta}_1=1/2, \bar{\theta}_2=-\pi/2$.

One may notice that the present eigenvalue configuration can be split into two pairs,
$(\zeta_1, \bar{\zeta}_1)$ and $(\zeta_2, \bar{\zeta}_2)$, with each pair corresponding to the eigenvalue configuration of a fundamental soliton. This invites the view that the present two-soliton should describe the nonlinear superposition of two fundamental solitons. Indeed, the solution graph in the top row of Fig. \ref{f:fig2} does more or less support this interpretation. However, this is not always the case. As we will see in the next section for the reverse-time NLS equation, even if a two-soliton's eigenvalue configuration can be split into groups of eigenvalues of fundamental solitons, this two-soliton may behave very differently from fundamental solitons and may not describe the nonlinear superposition of two fundamental solitons.

\vspace{0.2cm}
\noindent
(2) $\zeta_1, \zeta_2 \notin i\mathbb{R}_+$, and $\bar{\zeta}_1, \bar{\zeta}_2 \notin i\mathbb{R}_-$.

\vspace{0.1cm}
This is an interesting case, where all four eigenvalues are non-imaginary. Then due to the eigenvalue symmetry of $(\zeta, -\zeta^*)$, we must have $\zeta_2=-\zeta_1^*$, and $\bar{\zeta}_2=-\bar{\zeta}_1^*$, which make up an eigenvalue quartet. This eigenvalue configuration has been suggested in \cite{Gerdjikov2017}, but the corresponding solitons have not been studied. Unlike case (1) above, these eigenvalues cannot be split into groups of imaginary-eigenvalue pairs of fundamental solitons; thus they generate a different type of solitons. We find that these solitons always collapse repeatedly at pairs of spatial locations which are symmetric with respect to $x=0$. In addition, they move in two opposite directions as they collapse, which contrasts the fundamental soliton in Fig. \ref{f:fig1} and the stationary two-soliton in the top row of Fig. \ref{f:fig2}. To demonstrate, we take parameters as
\begin{eqnarray}
&& \hspace{-0.5cm}
\zeta_1=-\zeta_2^*=0.7+i, \quad \bar{\zeta}_1=-\bar{\zeta}_2^*=0.6-0.5i,  \label{2solitonRXpar2a} \\
&& \hspace{-0.5cm}
b_1=1+i, \quad \bar{b}_1=1-0.5i,   \label{2solitonRXpar2b}
\end{eqnarray}
and the corresponding solution is plotted in the middle row of Fig. \ref{f:fig2}. Notice that in addition to moving and collapsing, another interesting feature of this soliton is that its amplitudes also change as it moves. Specifically, the amplitude of the right-moving wave decreases exponentially with time, while the amplitude of the left-moving wave increases exponentially with time.

\vspace{0.2cm}
\noindent
(3) $\zeta_1, \zeta_2 \notin i\mathbb{R}_+$, and $\bar{\zeta}_1, \bar{\zeta}_2 \in i\mathbb{R}_-$.

\vspace{0.1cm}
This is an even more interesting configuration, where the two eigenvalues in the upper half plane are non-imaginary, but the two eigenvalues in the lower half plane are imaginary. Due to the eigenvalue symmetry of $(\zeta, -\zeta^*)$, the upper two non-imaginary eigenvalues must be related as $\zeta_2=-\zeta_1^*$. This eigenvalue configuration has not been mentioned or reported before. Here again, the four eigenvalues cannot be split into groups of imaginary-eigenvalue pairs of fundamental solitons, and they create a new type of two-solitons which differ from those in cases (1) and (2). To illustrate the dynamics of these new solitons, we choose parameter values
\begin{eqnarray}
&& \hspace{-0.8cm}
\zeta_1=-\zeta_2^*=0.3+1.2i, \quad \bar{\zeta}_1=-0.4i, \quad \bar{\zeta}_2=-i,   \label{2solitonRXpar3a}\\
&& \hspace{-0.8cm}
b_1=1+i, \quad \bar{\theta}_1=-\pi/4, \quad \bar{\theta}_2=-\pi.  \label{2solitonRXpar3b}
\end{eqnarray}
This eigenvalue configuration and the corresponding two-soliton are presented in the bottom row of Fig. \ref{f:fig2}. This soliton features two waves traveling in opposite directions, plus another stationary wave in the middle (at $x=0$). Both the traveling waves and the stationary wave collapse repeatedly over time. In addition, the amplitudes of the two traveling waves are changing, with the right-moving one decreasing with time and the left-moving one increasing with time. This two-soliton visually looks like a nonlinear superposition of a fundamental soliton as in Fig. \ref{f:fig1}, and a quartet-eigenvalue two-soliton as in the middle row of Fig. \ref{f:fig2}, even though its eigenvalue configuration does not suggest this visual appearance.

\vspace{0.2cm}
\noindent
(4) $\zeta_1, \zeta_2 \in i\mathbb{R}_+$, and $\bar{\zeta}_1, \bar{\zeta}_2 \notin i\mathbb{R}_-$.

\vspace{0.1cm}
The fourth eigenvalue configuration is the opposite of case (3), where the upper two $\mathbb{C}_+$ eigenvalues are purely imaginary, and the lower two $\mathbb{C}_-$ eigenvalues are non-imaginary. Due to the eigenvalue symmetry of $(\zeta, -\zeta^*)$, the lower two non-imaginary eigenvalues are related as $\bar{\zeta}_2=-\bar{\zeta}_1^*$. It is easy to check that solutions in this case can be linked to solutions of case (3). Specifically, suppose $q_4(x,t)$ is a solution of this case (4) with scattering data $S\equiv \{\zeta_k, \bar{\zeta}_k, v_{k0}, \bar{v}_{k0}, 1\le k\le 2\}$, and $q_3(x,t)$ is a solution with scattering data $S^*$. Since the conjugated eigenvalues
$\bar{\zeta}_1^*, \bar{\zeta}_2^*$ are in $\mathbb{C}_+$ and non-imaginary, and $\zeta_1^*, \zeta_2^*$ are in $\mathbb{C}_-$ and imaginary, the solution $q_3(x,t)$ then belongs to case (3). It is easy to recognize that $q_4(x,t)=-q_3^*(-x,-t)$. Thus, solution behaviors in case (4) can be inferred from those of case (3) without the need of separate analysis.

As we can see, two-solitons exhibit several new types of solutions which have not been seen before, and their dynamics cannot be understood from the dynamics of fundamental solitons. Three- and higher-solitons can be studied similarly, and additional novel behaviors can be expected.

\section{Dynamics of $N$-solitons in the reverse-time NLS equation}
To obtain general $N$-solitons in the reverse-time NLS equation (\ref{e:NLSRT}), we impose symmetry relations of discrete scattering data from Theorem \ref{Theorem2} in the general $N$-soliton formulae (\ref{Nsoliton}).  For a pair of discrete eigenvalues $(\zeta_k, \bar{\zeta}_k)$ with $\zeta_k\in \mathbb{C}_+$ and $\bar{\zeta}_k=-\zeta_k \in \mathbb{C}_-$,
without loss of generality we scale the eigenvector $v_{k0}$ of $\zeta_k$ so that $a_k=1$ in (\ref{vk0form}). Then using Theorem \ref{Theorem2}, we get
\[
v_{k0}=[1, b_k]^T, \quad \bar{v}_{k0}=[b_k, 1],
\]
where $b_k$ is a complex constant. This $N$-soliton has $2N$ free complex constants, $\{\zeta_k, b_k, 1\le k\le N\}$, with $\zeta_k\in \mathbb{C}_+$.

\begin{figure}[htbp]
\begin{center}
\includegraphics[width=0.5\textwidth]{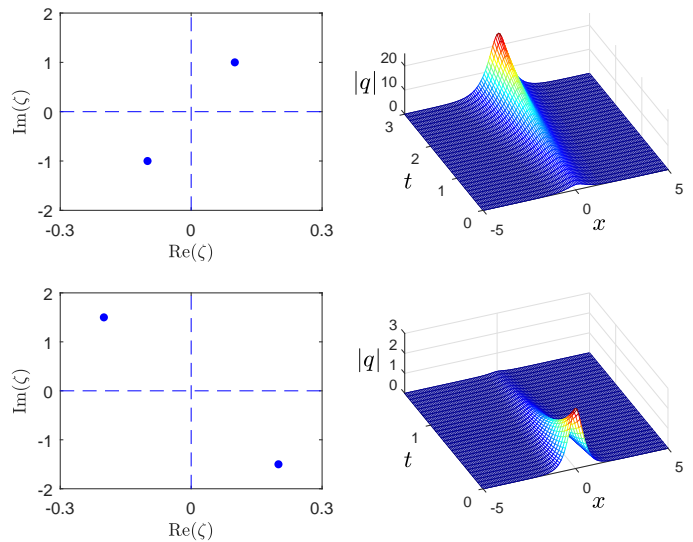}
\caption{Two fundamental solitons (\ref{NLSRT1}) in the reverse-time NLS equation (\ref{e:NLSRT}). The parameter values $(\zeta_1, b_1)$ are $(0.1+i, 1+0.5i)$ in the upper row and $(-0.2+1.5i, 1)$ in the lower row. Left column: eigenvalue configurations; right column: solution graphs.} \label{f:fig3}
\end{center}
\end{figure}

The fundamental soliton is obtained when we set $N=1$. In this case, simple algebra gives the analytical expression of this fundamental soliton as
\[ \label{NLSRT1}
q(x,t)=-4i\zeta_1b_1\frac{ e^{-4i\zeta_1^2t}}{e^{-2i\zeta_1x}+b_1^2e^{2i\zeta_1x}},
\]
where $\zeta_1$ and $b_1$ are free complex constants with $\zeta_1\in \mathbb{C}_+$.
This soliton does not move nor collapse. In addition, its amplitude grows or decays exponentially when $\zeta_1\notin i\mathbb{R}_+$ (it would grow/decay when $\zeta_1$ is in the first/second quadrant of the complex plane $\mathbb{C}$). For two sets of parameter values, $(\zeta_1, b_1)=(0.1+i, 1+0.5i)$ and $(-0.2+1.5i, 1)$, the graphs of this soliton are illustrated in Fig. \ref{f:fig3}.

\vspace{0.2cm}

\begin{figure}[htbp]
\begin{center}
\includegraphics[width=0.5\textwidth]{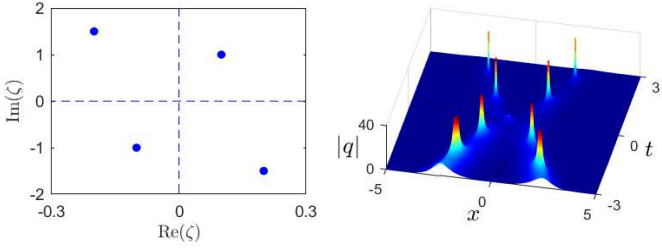}
\caption{A two-soliton in the reverse-time NLS equation (\ref{e:NLSRT}). The parameter values are given in Eq. (\ref{2solitonRTpar}).
Left panel: eigenvalue configuration; right panel: solution graph.} \label{f:fig4}
\end{center}
\end{figure}

Two-solitons can be obtained from (\ref{Nsoliton}) under the above scattering-data relations. Surprisingly, even though the fundamental solitons never collapse, the two-solitons would collapse repeatedly if $\zeta_1$ and $\zeta_2$ are not both purely imaginary. In addition, even though the fundamental solitons are stationary, the two-solitons would move in two opposite directions if $\zeta_1$ and $\zeta_2$ are not both purely imaginary. As an example, we choose parameter values as those in the two fundamental solitons of Fig. \ref{f:fig3}, i.e.,
\[ \label{2solitonRTpar}
\zeta_1=0.1+i, \hspace{0.2cm} \zeta_2=-0.2+1.5i, \hspace{0.2cm} b_1=1+0.5i, \hspace{0.2cm} b_2=1,
\]
and the corresponding two-soliton is plotted in Fig. \ref{f:fig4}. Its repeated collapsing and two-way motion can be seen. Meanwhile, along the directions of motion, the wave amplitudes also decrease over time.

One may notice that the eigenvalue configuration of this two-soliton can be split into pairs of eigenvalues of fundamental solitons in Fig. \ref{f:fig3}, but its dynamics is totally different from that of the fundamental solitons. Thus, dynamics of two-solitons in the reverse-time NLS equation (\ref{e:NLSRT}) is not a nonlinear superposition of two fundamental solitons and cannot be predicted from the dynamics of fundamental solitons. This is similar to two-solitons in the reverse-space NLS equation (\ref{e:NLSRX}), see Figs. \ref{f:fig1}-\ref{f:fig2}.

Even when both $\zeta_1, \zeta_2$ are purely imaginary, repeated collapsing can still occur for certain ranges of complex constants $b_1$ and $b_2$, although non-collapsing solutions also exist for the other ranges of $b_1$ and $b_2$ values. For instance, when $\zeta_1=i$ and $\zeta_2=1.5i$, we observed repeated collapsing for $b_1=1$ and $b_2=1+0.5i$ but non-collapsing for $b_1=b_2=1$.

\section{Dynamics of $N$-solitons in the reverse-space-time NLS equation}
To obtain general $N$-solitons in the reverse-space-time NLS equation (\ref{e:NLSRXT}), we impose symmetry relations of discrete scattering data from Theorem \ref{Theorem3} in the general $N$-soliton formulae (\ref{Nsoliton}). Specifically, we let
\[
v_{k0}=\left[1, \omega_k\right]^T, \quad  \bar{v}_{k0}=\left[1, \bar{\omega}_k\right],  
\]
where $\omega_k=\pm 1$, and $\bar{\omega}_k=\pm 1$. This $N$-soliton has $2N$ free complex constants, $\{\zeta_k, \bar{\zeta}_k, 1\le k\le N\}$, with $\zeta_k\in \mathbb{C}_+$ and $\bar{\zeta}_k\in \mathbb{C}_-$. In addition, each of $\omega_k$ and $\bar{\omega}_k$ has two choices between $\pm 1$.

\begin{figure}[htbp]
\begin{center}
\includegraphics[width=0.5\textwidth]{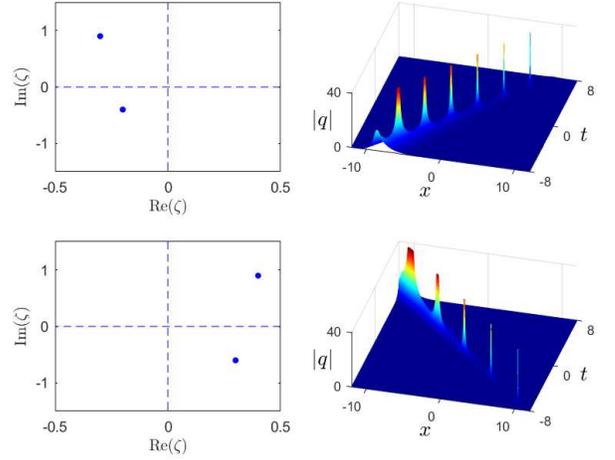}
\caption{Two fundamental solitons (\ref{NLSRXT1}) in the reverse-space-time NLS equation (\ref{e:NLSRXT}). The parameter values for the upper and lower rows are given in Eqs. (\ref{1solitonRXTpar1}) and (\ref{1solitonRXTpar2}) respectively. Left column: eigenvalue configurations; right column: solution graphs.} \label{f:fig5}
\end{center}
\end{figure}

For the fundamental soliton, we take $N=1$. In this case, the analytical expression of this fundamental soliton is
\[ \label{NLSRXT1}
q(x,t)=2i(\bar{\zeta}_1-\zeta_1)\frac{\bar{\omega}_1 e^{-2i\bar{\zeta}_1x-4i\bar{\zeta}_1^2t}}{1+\omega_1\bar{\omega}_1 e^{-2i(\bar{\zeta}_1-\zeta_1)x-4i(\bar{\zeta}_1^2-\zeta_1^2)t}},
\]
where $\omega_1=\pm 1$, $\bar{\omega}_1=\pm 1$, and $\zeta_1\in \mathbb{C}_+$, $\bar{\zeta}_1\in \mathbb{C}_-$ are free. This soliton moves at velocity $V=-2\mbox{Im}(\bar{\zeta}_1^2-\zeta_1^2)/\mbox{Im}(\bar{\zeta}_1-\zeta_1)$. On the line $x=Vt$, its amplitude $|q|$ changes as
\[ \label{XTqcenter}
|q(t)|=2|\bar{\zeta}_1-\zeta_1| \frac{e^{\beta t}}{1+\omega_1\bar{\omega}_1 e^{i\gamma t}},
\]
where
\begin{eqnarray}
&& \beta=2V\mbox{Im}(\bar{\zeta}_1)+4\mbox{Im}(\bar{\zeta}_1^2), \\
&& \gamma=-2V\mbox{Re}(\bar{\zeta}_1-\zeta_1)-4\mbox{Re}(\bar{\zeta}_1^2-\zeta_1^2).
\end{eqnarray}
Thus, this soliton's amplitude is growing or decaying exponentially (depending on the sign of $\beta$). In addition, it periodically collapses with period
%$\pi\mbox{Im}(\bar{\zeta}_1-\zeta_1)/[2|\bar{\zeta}_1-\zeta_1|^2\mbox{Im}(\bar{\zeta}_1+\zeta_1)]$.
$2\pi/|\gamma|$ if $\gamma\ne 0$, or equivalently, if $\mbox{Im}(\zeta_1+\bar{\zeta}_1)\ne 0$, because it is easy to show that $\gamma$ can be rewritten as $\gamma=4|\zeta_1-\bar{\zeta}_1|^2\mbox{Im}(\zeta_1+\bar{\zeta}_1)/\mbox{Im}(\bar{\zeta}_1-\zeta_1)$.
%A unique feature of this soliton is that it does not have free position and phase parameters, which is caused by the fact that the %reverse-space-time NLS equation (\ref{e:NLSRXT}) does not admit spatial-translation and temporal-translation invariances.
For the two sets of parameters
\[
\zeta_1=-0.3+0.9i, \quad \bar{\zeta}_1=-0.2-0.4i, \quad \omega_1=\bar{\omega}_1=1,        \label{1solitonRXTpar1}
\]
and
\[
\zeta_1=0.4+0.9i, \hspace{0.25cm} \bar{\zeta}_1=0.3-0.6i, \hspace{0.25cm} \omega_1=1, \hspace{0.25cm} \bar{\omega}_1=-1,        \label{1solitonRXTpar2}
\]
graphs of the two fundamental solitons are displayed in the upper and lower rows of Fig. \ref{f:fig5} respectively. In the former case, the soliton moves at velocity $V\approx 1.0769$ (to the right). Along the line $x=Vt$, $|q|$ decreases exponentially at the rate of $e^{\beta t}$ with $\beta\approx -0.2215$, and collapses repeatedly with collapsing period $2\pi/|\gamma|\approx 2.4024$. In the latter case, the soliton moves at velocity $V= -1.44$ (to the left). Along the line $x=Vt$, $|q|$ increases exponentially at the rate of $e^{\beta t}$ with $\beta=0.288$, and collapses repeatedly with collapsing period $2\pi/|\gamma|\approx 3.4752$.

\begin{figure}[htbp]
\begin{center}
\includegraphics[width=0.5\textwidth]{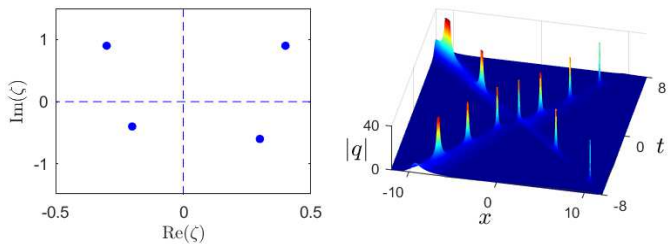}
\caption{A two-soliton in the reverse-space-time NLS equation (\ref{e:NLSRXT}). The parameter values are given in Eqs. (\ref{2solitonRXTpara})-(\ref{2solitonRXTparb}). Left panel: eigenvalue configuration; right panel: solution graph.}  \label{f:fig6}
\end{center}
\end{figure}

If we take $N=2$ in the formula (\ref{Nsoliton}), we get two-soliton solutions. Using the same parameters as in the two fundamental solitons of Fig. \ref{f:fig5}, i.e.,
\begin{eqnarray}
&& \hspace{-1.2cm}
\zeta_1=-0.3+0.9i, \hspace{0.1cm} \zeta_2=0.4+0.9i, \hspace{0.1cm} \bar{\zeta}_1=-0.2-0.4i,    \label{2solitonRXTpara}  \\
&& \hspace{-1.2cm} \bar{\zeta}_2=0.3-0.6i, \hspace{0.1cm} \omega_1=\omega_2=\bar{\omega}_1=1, \hspace{0.1cm} \bar{\omega}_2=-1,     \label{2solitonRXTparb}
\end{eqnarray}
the corresponding two-soliton is plotted in Fig. \ref{f:fig6}. This two-soliton moves in two directions, and collapses repeatedly as it moves. The amplitudes of the two moving waves change with time as well. When compared to the two fundamental solitons in Fig. \ref{f:fig5}, we see that this two-soliton does describe the nonlinear superposition between those two fundamental solitons, as the eigenvalue configuration of this two-soliton suggests (this eigenvalue configuration can be split into two groups corresponding to the eigenvalue configurations of the two fundamental solitons in Fig. \ref{f:fig5}). Thus, the reverse-space-time NLS equation (\ref{e:NLSRXT}) is the only nonlocal equation in this paper where a two-soliton is indeed a nonlinear superposition of two fundamental solitons.
%This behavior holds for higher solitons as well.

\section{Summary and discussion}

In this article, we have derived general $N$-solitons in the reverse-space, reverse-time, and reverse-space-time nonlinear Schr\"odinger equations (\ref{e:NLSRX})-(\ref{e:NLSRXT}) from the Riemann-Hilbert solutions of the AKNS hierarchy. We have shown that symmetry relations of the scattering data in these nonlocal equations differ greatly from those of the local NLS equation, which lead to dramatically different solution behaviors in these nonlocal equations. We have found that a generic feature of solutions in these nonlocal equations
is repeated collapsing. In addition, multi-solitons often do not describe a nonlinear superposition of fundamental solitons, and they exhibit distinctive solution patterns which have not been seen before. These findings reveal the novel and rich soliton structures in the three nonlocal NLS equations (\ref{e:NLSRX})-(\ref{e:NLSRXT}), and they invite further investigations of solitons and multi-solitons in the other nonlocal equations.

The new symmetry properties of scattering data in these nonlocal equations also help resolve some open questions left over in previous Riemann-Hilbert derivations of solitons. In that treatment, it was always assumed that the numbers of eigenvalues (known as zeros of the Riemann-Hilbert problem) in the upper and lower complex planes, counting multiplicity, were equal to each other \cite{Zakharov1984,Yang2010,YangJMP2003}.
%The reason is that only in that case can one pair up those $\mathbb{C}_+$ and $\mathbb{C}_-$ eigenvalues and build a sequence of constituent solution matrices with those eigenvalues as zeros or poles in the lower or upper complex planes \cite{Zakharov1984,Yang2010,YangJMP2003}. For instance, in the present Riemann-Hilbert solution (\ref{qsolution})-(\ref{rsolution}), we had assumed that the numbers of eigenvalues in $\mathbb{C}_+$ and $\mathbb{C}_-$ were both equal to $N$ and were all simple \cite{Yang2010}.
A natural open question was: what would happen if the numbers of eigenvalues in $\mathbb{C}_+$ and $\mathbb{C}_-$ are not equal to each other? This question could not be addressed in the local (focusing) NLS equation, because in that case, eigenvalues always appear as conjugate pairs and thus always come in equal numbers in $\mathbb{C}_+$ and $\mathbb{C}_-$ \cite{Ablowitz1981,Zakharov1984,Faddeev1987,Ablowitz1991,Yang2010}. However, in the reverse-space NLS equation (\ref{e:NLSRX}) and reverse-space-time NLS equation (\ref{e:NLSRXT}), eigenvalues in $\mathbb{C}_+$ and $\mathbb{C}_-$ are totally independent (see Theorems \ref{Theorem1} and \ref{Theorem3}). As a consequence, it is now possible for eigenvalues in $\mathbb{C}_+$ and $\mathbb{C}_-$ to appear in unequal numbers. To address this question, let us consider the simplest case in the reverse-space NLS equation (\ref{e:NLSRX}), where there is a single pair of purely imaginary eigenvalues $(\zeta_1, \bar{\zeta}_1)$, but now both in $\mathbb{C}_+$. It is easy to verify that the previous fundamental soliton (\ref{1solitonRX}) still satisfies the reverse-space NLS equation (\ref{e:NLSRX}) even though both $\zeta_1$ and $\bar{\zeta}_1$ are in $i\mathbb{R}_+$. But in this case, this ``fundamental soliton" is unbounded in space for all times, because it grows exponentially in either the positive or negative $x$ directions. This example tells us that, when the Riemann-Hilbert problem has unequal numbers of zeros (eigenvalues) in the upper and lower complex planes, it would produce solutions which are unbounded in space (thus never solitons).

%However, when the numbers of eigenvalues in $\mathbb{C}_+$ and $\mathbb{C}_-$ are equal to each other (as previously always assumed), the solutions out of the Riemann-Hilbert method will always be localized in space and thus always be soliton solutions, as this article has shown for the three nonlocal equations (\ref{e:NLSRX})-(\ref{e:NLSRXT}). For the derivation of soliton solutions, this is an advantage, because in the Darboux transformation and bilinear approaches as used earlier for some nonlocal equations \cite{WYY2016,HXLM2016,Stalin2017wrong,Zhang1,Peckan2017arxiv}, it was not clear how to constrain parameters to obtain general multi-soliton solutions.

\section*{Acknowledgment}
This material is based upon work supported by the Air Force Office of Scientific Research under award number FA9550-12-1-0244, and the National Science Foundation under award number DMS-1616122.


\begin{thebibliography}{10}
\bibitem{Ablowitz1981}
M.J. Ablowitz and H. Segur, \emph{Solitons and Inverse Scattering Transform} (SIAM, Philadelphia, 1981).

\bibitem{Zakharov1984}
S.P. Novikov, S.V. Manakov, L.P. Pitaevskii and V.E. Zakharov, \emph{Theory of Solitons} (Plenum, New York, 1984).

\bibitem{Faddeev1987} L. Takhtadjan and L. Faddeev, \emph{The Hamiltonian Approach to Soliton Theory} (Springer Verlag, Berlin, 1987).

\bibitem{Ablowitz1991} M.J. Ablowitz and P.A. Clarkson, \emph{Solitons, Nonlinear Evolution Equations and Inverse
Scattering} (Cambridge University Press, 1991).

\bibitem{Yang2010} J. Yang, \emph{Nonlinear Waves in Integrable and Non integrable Systems} (SIAM, Philadelphia, 2010).

\bibitem{AblowitzMussPRL2013}
M.J. Ablowitz and Z.H. Musslimani,
``Integrable nonlocal nonlinear Schr\"odinger equation",
Phys. Rev. Lett. 110, 064105 (2013).

\bibitem{PTNLSmagnetics} T.A. Gadzhimuradov and A.M. Agalarov,
``Towards a gauge-equivalent magnetic structure of the nonlocal nonlinear Schr\"odinger equation",
Phys. Rev. A 93, 062124 (2016).

\bibitem{Yang_review}
V.V. Konotop, J. Yang and D.A. Zezyulin,
``Nonlinear waves in \PT-symmetric systems,"
Rev. Mod. Phys. 88, 035002 (2016).

\bibitem{AblowitzMussNonli2016}
M.J. Ablowitz and Z.H. Musslimani,
``Inverse scattering transform for the integrable nonlocal nonlinear Schrodinger equation,"
Nonlinearity 29, 915-946 (2016).

\bibitem{WYY2016}
X.\ Y.\ Wen, Z.\ Yan and Y.\ Yang,
``Dynamics of higher-order rational solitons for the nonlocal nonlinear Schrodinger equation with the self-induced parity-time-symmetric potential",
Chaos 26, 063123 (2016).

\bibitem{HXLM2016}
X.\ Huang and L.\ M.\ Ling,
``Soliton solutions for the nonlocal nonlinear Schrodinger equation,"
Eur. Phys. J. Plus 131, 148 (2016).

\bibitem{Gerdjikov2017} V.S. Gerdjikov and A. Saxena,
``Complete integrability of nonlocal nonlinear Schr\"odinger equation",
J. Math. Phys. 58, 013502 (2017).

\bibitem{Stalin2017wrong}
S. Stalin, M. Senthilvelan, and M. Lakshmanan,
``Nonstandard bilinearization of \PT-invariant nonlocal Schr\"odinger equation: bright soliton solutions",
Phys. Lett. A 381, 2380-2385 (2017).

\bibitem{Caudrelier}
V. Caudrelier,
``Interplay between the inverse scattering method and Fokas' unified transform with an application,"
Stud. Appl. Math. DOI: 10.1111/sapm.12190 (2017).

\bibitem{Zhang1} K. Chen and D.J. Zhang,
``Solutions of the nonlocal nonlinear Schr\"odinger hierarchy via reduction",
Appl. Math. Lett., 75, 82-88 (2018).

\bibitem{Peckan2017arxiv}
M. G\"urses and A. Pekcan, ``Nonlocal nonlinear Schr\"odinger equations and their soliton solutions",
arXiv:1707.07610 [nlin.SI] (2017).

\bibitem{BYJYrogue} B.\ Yang  and J.\ Yang, ``General rogue waves in the nonlocal \PT-symmetric
nonlinear Schr\"odinger equation", 	arXiv:1711.05930 [nlin.SI] (2017).

\bibitem{AblowitzMussPRE2014}
M.J. Ablowitz and Z.H. Musslimani,
``Integrable discrete \PT-symmetric model",
Phys. Rev. E 90, 032912 (2014).

\bibitem{Yan} Z. Yan,
``Integrable \PT-symmetric local and nonlocal vector nonlinear Schroinger equations: A unified two-parameter model,"
Appl. Math. Lett. 47, 61-68 (2015).

\bibitem{Khara2015}
A. Khara and A. Saxena,
``Periodic and hyperbolic soliton solutions of a number of nonlocal nonlinear equations",
J. Math. Phys. 56, 032104 (2015).

%\bibitem{Zhu1} C.Q. Song, D.M. Xiao and Z.N. Zhu, ``A general integrable nonlocal coupled nonlinear Schr\"odinger equation",
%arXiv:1505.05311 [nlin.SI] (2015).

\bibitem{Fokas2016} A.S. Fokas,
``Integrable multidimensional versions of the nonlocal nonlinear Schr\"odinger equation",
Nonlinearity 29, 319-324 (2016).

\bibitem{GerdjikovNwave2016}
V.S. Gerdjikov, G.G. Grahovski, and R. I. Ivanov,
``The $N$-wave equations with \PT symmetry",
Theoret. Math. Phys. 188, 1305-1321 (2016).

\bibitem{Chow} Z. X. Xu, and K. W. Chow,
``Breathers and rogue waves for a third order nonlocal partial differential equation by a bilinear transformation",
Appl. Math. Lett. 56, 72-77 (2016).

\bibitem{JZN2017}
J.\ L.\ Ji  and Z.\ N.\ Zhu,
``On a nonlocal modified Korteweg-de Vries equation: Integrability, Darboux transformation and soliton solutions,"
Commun. Nonlinear Sci. Numer. Simul. 42, 699-708 (2017).

\bibitem{Lou2}
S.Y. Lou and F. Huang,
``Alice-Bob physics: coherent solutions of nonlocal KdV systems",
Scientific Reports 7, 869 (2017).

\bibitem{AblowitzMussSAPM}
M.J. Ablowitz and Z.H. Musslimani,
``Integrable nonlocal nonlinear equations",
Stud. Appl. Math. 139, 7-59 (2017).

\bibitem{ZhoudNLS} Z.X. Zhou, ``Darboux transformations and global solutions for a nonlocal derivative nonlinear Schr\"odinger equation",
arXiv:1612.04892 [nlin.SI] (2016).

%\bibitem{ZhouDS} Z.X. Zhou, ``Darboux transformations and global explicit solutions for nonlocal Davey-Stewartson I equation",
%arXiv:1612.05689 [nlin.SI] (2016).

\bibitem{Zhu2} J.L. Ji and Z.N. Zhu,
``Soliton solutions of an integrable nonlocal modified Korteweg-de Vries equation through inverse scattering transform",
J. Math. Anal. Appl. 453, 973-984 (2017).

\bibitem{HePPTDS} J. G. Rao, Y. Cheng and J.S. He,
``Rational and semi-rational solutions of the nonlocal Davey-Stewartson equations",
Stud. Appl. Math. 139, 568-598 (2017).

\bibitem{Zhu3} L.Y. Ma, S.F. Shen and Z.N. Zhu,
``Soliton solution and gauge equivalence for an integrable nonlocal complex modified Korteweg-de Vries equation",
J. Math. Phys. 58, 103501 (2017).

\bibitem{BYJY2017}
B.\ Yang  and J.\ Yang, ``Transformations between nonlocal and local integrable equations,"  Stud. Appl. Math. DOI: 10.1111/sapm.12195 (2017).

\bibitem{nonlocalFordy2017}
M. G\"urses,
``Nonlocal Fordy-Kulish equations on symmetric spaces",
Phys. Lett. A 381, 1791-1794 (2017).

\bibitem{AblowitzSG2017}
M.J. Ablowitz, B.F. Feng, X.D. Luo and Z.H. Musslimani,
``Inverse scattering transform for the nonlocal reverse space-time Sine-Gordon, Sinh-Gordon and nonlinear Schr\"odinger equations with nonzero boundary conditions", arXiv:1703.02226 [math-ph] (2017).

\bibitem{BYnonlocalDS}
B. Yang and Y. Chen, ``Dynamics of rogue waves in the partially \PT-symmetric nonlocal Davey-Stewartson systems", arXiv:1710.07061 [math-ph] (2017).

\bibitem{Zhang2} K. Chen, X. Deng, S.Y. Lou, D.J. Zhang, ``Solutions of local and nonlocal equations reduced from the AKNS hierarchy",
arXiv:1710.10479 [nlin.SI] (2017).

\bibitem{ZS1972}
V.E. Zakharov and A.B. Shabat,
``Exact theory of two-dimensional self-focusing and one-dimensional self-modulation of waves in nonlinear media",
Zh. E'ksp. Teor. Fiz. 61, 118 (1971) [Sov. Phys. JETP 34, 62 (1972)].

\bibitem{AKNS1974}
M.J. Ablowitz, D.J. Kaup, A.C. Newell and H. Segur,
``The inverse scattering transform --- Fourier analysis for nonlinear problems",
Stud. Appl. Math. 53, 249 (1974).

\bibitem{ZS1979}
V.E. Zakharov and A.B. Shabat,
``Integration of the nonlinear equations of mathematical physics by the method of the inverse scattering problem II",
Funk. Anal. Prilozh. 13, 13-22 (1979) [Funct. Anal. Appl. 13, 166-174 (1979)].

\bibitem{YangJMP2003}
V.S. Shchesnovich and J. Yang,
``General soliton matrices in the Riemann-Hilbert problem for integrable nonlinear equations."
J. Math. Phys. 44, 4604-4639 (2003).

\end{thebibliography}
\end{document}